\documentclass[aps,pra,showpacs,twocolumn,groupedaddress]{revtex4}
\usepackage{graphicx,amsmath,amssymb}
\begin{document}
\newtheorem{Theorem}{Theorem}

\title{A relation between fidelity and quantum adiabatic evolution}

\author{Zhaohui Wei and Mingsheng Ying}

\affiliation{ State Key Laboratory of Intelligent Technology and
Systems, Department of Computer Science and Technology, Tsinghua
University, Beijing, China, 100084}

\begin{abstract}

Recently, some quantum algorithms have been implemented by quantum
adiabatic evolutions. In this paper, we discuss the accurate
relation between the running time and the distance of the initial
state and the final state of a kind of quantum adiabatic evolutions.
We show that this relation can be generalized to the case of mixed
states.

\end{abstract}

\maketitle

Implementing quantum algorithms via quantum adiabatic evolutions is
a novel paradigm for the design of quantum algorithms, which was
proposed by Farhi et al. \cite{FGGS00}. In a quantum adiabatic
algorithm, the evolution of the quantum register is governed by a
hamiltonian that varies continuously and slowly. At the beginning,
the state of the system is the ground state of the initial
hamiltonian. If we encode the solution of the algorithm in the
ground state of the final hamiltonian and if the hamiltonian of the
system evolves slowly enough, the quantum adiabatic theorem
guarantees that the final state of the system will differ from the
ground state of the final hamiltonian by a negligible amount. Thus
after the quantum adiabatic evolution we can get the solution with
high probability by measuring the final state. For example, Quantum
search algorithm proposed by Grover \cite{GROVER97} has been
implemented by quantum adiabatic evolution in \cite{RC02}. Recently,
the new paradigm for quantum computation has been tried to solve
some other interesting and important problems
\cite{TH03,TDK01,FGG01}. For example, T. D. Kieu has proposed a
quantum adiabatic algorithm for Hilbert's tenth problem \cite{TDK01}
, while this problem is known to be mathematically noncomputable.

Usually, after the design of a quantum adiabatic evolution, the
estimation of the running time is not easy. In \cite{RC02}, Roland
et al. introduced a policy to design a class of quantum local
adiabatic evolutions with a performance that can be estimated
accurately. Using this policy Roland et al. reproduced quantum
search algorithm, which is as good as Grover's algorithm.

For convenience of the readers, we briefly recall the local
adiabatic algorithm. Suppose $H_0$ and $H_T$ are the initial and the
final Hamiltonians of the system, we choose them as
\begin{equation}
H_0=I-|\alpha\rangle\langle\alpha|,
\end{equation}
and
\begin{equation}
H_T=I-|\beta\rangle\langle\beta|,
\end{equation}
where $|\alpha\rangle$ is the initial state of the system and
$|\beta\rangle$ is the final state that encodes the solution. Then
we let the system vary under the following time dependent
Hamiltonian:
\begin{equation}
H(t)=(1-s)H_0+sH_T,
\end{equation}
where $s=s(t)$ is a monotonic function with $s(0)=0 $ and $s(T)=1$
($T$ is the running time of the evolution). Let $|E_0,t\rangle$ and
$|E_1,t\rangle$ be the ground state and the first excited state of
the Hamiltonian at time t, and let $E_0(t)$ and $E_1(t)$ be the
corresponding eigenvalues. The adiabatic theorem \cite{LIS55} shows
that we have
\begin{equation}
|\langle E_0,T|\psi(T)\rangle|^{2}\geq1-\varepsilon^2,
\end{equation}
provided that
\begin{equation}
\frac{D_{max}}{g_{min}^2}\leq\varepsilon,\ \ \ \
0<\varepsilon\ll1,
\end{equation}
where $g_{min}$ is the minimum gap between $E_0(t)$ and $E_1(t)$
\begin{equation} g_{min}=\min_{0\leq t \leq T}[E_1(t)-E_0(t)],
\end{equation}
and $D_{max}$ is a measurement of the evolving rate of the
Hamiltonian
\begin{equation}
D_{max}=\max_{0\leq t \leq
T}|\langle\frac{dH}{dt}\rangle_{1,0}|=\max_{0\leq t \leq
T}|\langle E_1,t|\frac{dH}{dt}|E_0,t\rangle|.
\end{equation}

In the local adiabatic evolution of \cite{RC02},
\begin{equation}
|\alpha\rangle=\frac{1}{\sqrt{N}}\sum\limits_{i=1}^{N}{|i\rangle}, \
|\beta\rangle=|m\rangle ,
\end{equation}
where $N$ is the size of the database and $m$ is the solution of the
search problem. To evaluate the running time of the adiabatic
evolution, Roland and Cerf calculated accurately the gap $g_{min}$
in Eq. (2) and just estimated the quantity $D_{max}$ in Eq. (3)
using the bound
\begin{equation}
|\langle\frac{dH}{dt}\rangle_{1,0}|\leq |\frac{ds}{dt}|.
\end{equation}
To evaluate the performance of this algorithms, this is enough,
because calculating accurately the quantity $D_{max}$ in (7) can't
improve the result much. However, in this paper we will take into
account all the related quantities. Later we will find that this
will result in a simple and intrinsical relation between the running
time of the adiabatic evolution and the distance of the initial and
the final states.

In this paper, we will choose fidelity, one of the most popular
distance measures in the literature, as the measure of the hardness
to evolve from one state to another using adiabatic evolutions.

The fidelity of states $\rho$ and $\sigma$ is defined to be
\begin{equation}
F(\rho,\sigma)=tr\sqrt{\rho^{1/2}\sigma\rho^{1/2}}.
\end{equation}
Although fidelity is not a metric, its modified version
\begin{equation}
A(\rho,\sigma)=\arccos{F(\rho,\sigma)}
\end{equation}
is easily proved to be a metric \cite{Nielsen00}. Another important
metric for the distance between quantum states we will use in this
paper is the trace distance defined as
\begin{equation}
D(\rho,\sigma)=\frac{1}{2}tr|\rho-\sigma|.
\end{equation}
Now, we can represent the main result as the following theorem.

\begin{Theorem}Suppose $|\alpha\rangle$ and $|\beta\rangle$ are two states
of a quantum system. We can make the system evolve from the initial
state $|\alpha\rangle$ to the final state $|\beta\rangle$ by a
quantum adiabatic evolution, if we set the initial Hamiltonian $H_0$
and the final Hamiltonian $H_T$ of the adiabatic evolution as
follows:
$$H_0=I-|\alpha\rangle\langle\alpha|,$$
$$H_T=I-|\beta\rangle\langle\beta|.$$
To success with a probability at least $1-\varepsilon^2$, the
minimal running time that the adiabatic evolution requires is
\begin{equation}
T(|\alpha\rangle,|\beta\rangle)=\frac{1}{\varepsilon}\cdot\tan{(\arccos{F(|\alpha\rangle,|\beta\rangle)})},
\end{equation}
where
\begin{equation}
F(|\alpha\rangle,|\beta\rangle)=|\langle\alpha|\beta\rangle|
\end{equation}
is the fidelity between $|\alpha\rangle$ and $|\beta\rangle$.
\end{Theorem}

{\it Proof.} Let
$$H(s)=(1-s)(I-|\alpha\rangle\langle\alpha|)+s(I-|\beta\rangle\langle\beta|),$$
where $s=s(t)$ is a function of $t$ as described above.

It is not easy to calculate the eigenvalues of $H(s)$ in the
computational basis. We use the following orthonormal basis
$\{|i\rangle, 1\leq i\leq N\}$ to eliminate the difficulty:
\begin{equation}
|1\rangle=|\alpha\rangle,
\end{equation}
\begin{equation}
|2\rangle=\frac{1}{c}(|\beta\rangle-\langle\alpha|\beta\rangle|\alpha\rangle),
\end{equation}
where
$c=||\beta\rangle-\langle\alpha|\beta\rangle|\alpha\rangle|=\sqrt{1-|\langle\alpha|\beta\rangle|^{2}}$.
We don't need to care about $|\alpha_i\rangle$ for $i=3,4,...,N$.
Then we have
\begin{equation}
|\beta\rangle=c|2\rangle+\langle\alpha|\beta\rangle|1\rangle.
\end{equation}
Now it is not difficult to check that, in the new orthonormal basis,
$H(s)$ has a form of
\begin{equation}
H(s)=
\begin{pmatrix}
-s|\langle\alpha|\beta\rangle|^2+s & -sc\langle\alpha|\beta\rangle \\
-sc\langle\alpha|\beta\rangle^{*} & -sc^2+1\\
& & I_{(N-2)\times(N-2)}\\
\end{pmatrix},
\end{equation}
where the empty spaces of the matrix are all zeroes. Letting
$a=|\langle \alpha|\beta\rangle|$, it is easy to get the two lowest
eigenvalues of $H(s)$
\begin{equation}
E_i(t)=\frac{1}{2}(1\pm\sqrt{1-4(1-a^2)s(1-s)}), \ i=0,1,
\end{equation} and two corresponding eigenvectors
\begin{equation}
|E_i,t\rangle=\frac{1}{\sqrt{1+y_i^2}}(|1\rangle+y_i|2\rangle), \
i=0,1,
\end{equation}
where
\begin{equation}
y_i=\frac{\sqrt{1-a^2}}{a}-\frac{E_i(t)}{sa\sqrt{1-a^2}}  \
 \ (s\neq0).
\end{equation} Thus, we get $g(s)$:
\begin{equation}
g(s)=\sqrt{1-4(1-a^2)s(1-s)}.
\end{equation}
On the other hand, it is easy to kown
\begin{equation}
\frac{dH}{ds}=H_T-H_0=\frac{H(s)-H_0}{s} \ \ (s\neq0).
\end{equation}
Because $|E_0,t\rangle$ and $|E_1,t\rangle$ are
eigenvectors of $H(s)$, we have
\begin{equation}
\langle E_0,t|E_1,t\rangle = 0,
\end{equation}
and
\begin{equation}
\langle E_0,t|H(s)|E_1,t\rangle = 0.
\end{equation}
Then it can be shown that
\begin{equation}
|\langle\frac{dH}{ds}\rangle_{0,1}|=|\langle
E_0,t|\frac{H(s)-H_0}{s}|E_1,t\rangle|=|\frac{\langle
E_0,t|1\rangle\langle 1|E_1,t\rangle}{s}|.
\end{equation}
So
\begin{equation}
|\langle\frac{dH}{dt}\rangle_{0,1}|=|\frac{ds}{dt}|\cdot|\langle\frac{dH}{ds}\rangle_{0,1}|=|\frac{ds}{dt}|\cdot\frac{1}{s\sqrt{(1+y_0^2)(1+y_1^2)}}.
\end{equation}
Substituting Eq.(21) into Eq.(27) we have
\begin{equation}
|\langle\frac{dH}{dt}\rangle_{0,1}|=|\frac{ds}{dt}|\cdot\frac{a\sqrt{1-a^2}}{\sqrt{1-4(1-a^2)s(1-s)}}.
\end{equation}
In a local adiabatic evolution \cite{RC02}, the adiabaticity
condition (5) must be satisfied at any instant of time t,
\begin{equation}
|\frac{ds}{dt}|\cdot\frac{a\sqrt{1-a^2}}{\sqrt{1-4(1-a^2)s(1-s)}}\leq
\varepsilon(1-4(1-a^2)s(1-s)).
\end{equation}
To make the evolution as fast as possible, we can let $s(t)$ satisfy
the equation
\begin{equation}
\frac{ds}{dt}=\varepsilon\frac{(1-4(1-a^2)s(1-s))^{\frac{3}{2}}}{a\sqrt{1-a^2}}.
\end{equation}
By integration, we can get the lower bound of the running time of
the whole evolution
\begin{equation}
T(|\alpha\rangle,
|\beta\rangle)=\frac{1}{\varepsilon}\cdot\frac{\sqrt{1-a^2}}{a}=\frac{1}{\varepsilon}\cdot\tan{(\arccos{F(|\alpha\rangle,|\beta\rangle)})}.
\end{equation}

That completes the proof of this theorem. \hfill $\Box$

In fact, it is interesting to notice that we can rewrite the
relation above as
\begin{equation}
T(|\alpha\rangle,
|\beta\rangle)=\frac{1}{\varepsilon}\cdot\frac{D(|\alpha\rangle,|\beta\rangle)}{F(|\alpha\rangle,|\beta\rangle)},
\end{equation}
where $D(|\alpha\rangle,|\beta\rangle)$ is the trace distance
between $|\alpha\rangle$ and $|\beta\rangle$.

In \cite{RC02}, the fidelity between the initial state and the final
state of the local quantum adiabatic evolution is
$\frac{1}{\sqrt{N}}$. According to Theorem 1 the running time is
$O(\sqrt{N})$. This is consistent with the result of \cite{RC02}.

Similarly, In \cite{DKK02} S. Das et al. implement Deutsch's
algorithm \cite{DD85,DJ92} by an adiabatic evolution of the form
discussed in Theorem 1. In that work, if the system has $n$ qubits,
$|\alpha\rangle$ and $|\beta\rangle$ will be $(N=2^n)$
\begin{equation}
|\alpha\rangle=\frac{1}{\sqrt{N}}\sum\limits_{i=0}^{N-1}{|i\rangle},
\end{equation}
\begin{equation}
|\beta\rangle=\mu|0\rangle+\frac{\nu}{\sqrt{N-1}}\sum\limits_{i=1}^{N-1}{|k\rangle},
\end{equation}
with
\begin{equation}
\mu=\frac{1}{N}|\sum\limits_{x\in\{0,1\}^n}^{}{(-1)^{f(x)}}|,
\end{equation}
\begin{equation}
\nu=1-\mu.
\end{equation}
Here, the function $f:\{0,1\}^n\rightarrow\{0,1\}$ is either
constant (i.e., all outputs are identical) or balanced (i.e., has an
equal number of 0's and 1's as outputs), and our task is to decide
whether it is constant or not. It is not difficult to know that in
this case
\begin{equation}
F(|\alpha\rangle,|\beta\rangle) = |\langle\alpha|\beta\rangle| =
\frac{1}{\sqrt{N}} \ or \ \sqrt{1-\frac{1}{N}}.
\end{equation}
To make the algorithm success, we must let the running time of the
adiabatic evolution be long enough. So the minimal running time
should be $O(\sqrt{N})$. This result is consistent with
\cite{DKK02}.

Let's try to explain the meaning of the theorem. As we know,
$\arccos (F(\rho,\sigma))$ measures the distance of two quantum
states $\rho$ and $\sigma$ \cite{Nielsen00}. Quantum adiabatic
evolution, On the other hand, changes the state of a quantum system
from the initial state $|\alpha\rangle$ to the final state
$|\beta\rangle$. Our theorem says that if the precision of the
evolution is fixed, the minimal running time
$T(|\alpha\rangle,|\beta\rangle)$ will be direct proportional to the
tangent of $\arccos (F(|\alpha\rangle,|\beta\rangle))$. The smaller
the distance of the two states is, the shorter the running time of
the adiabatic evolution will be. This is consistent with our
intuition. However, we should notice that as the fidelity becomes
smaller, the running time will increase very quickly. Fore example,
when $F(|\alpha\rangle,|\beta\rangle)$ is 0.5, the running time
$T(|\alpha\rangle,|\beta\rangle)$ is $\frac{\sqrt{3}}{\varepsilon}$.
While as $F(|\alpha\rangle,|\beta\rangle)$ tends to 0, the running
time tends to infinite.

In the quantum adiabatic evolution, the initial and the final states
are pure. Using Uhlmann's theorem \cite{Nielsen00} we can generalize
the relation to the case of mixed states. Suppose $\rho$ and
$\sigma$ are two states of a quantum system $A$. Let $B$ is another
system and $A$ is a part of $B$. Suppose an adiabatic evolution
makes the state of $B$ evolve from $|\psi\rangle$ to
$|\varphi\rangle$ and in the same evolution the state of $A$ evolves
from $\rho$ to $\sigma$. We may ask --- is there any relation
between the running time of the adiabatic evolution and the fidelity
of $\rho$ and $\sigma$? We say yes by the following theorem.

\begin{Theorem}
Suppose $\rho$ and $\sigma$ are two mixed states, and let
\begin{equation}
T(\rho,\sigma)=\min_{|\psi\rangle,|\varphi\rangle}T(|\psi\rangle,|\varphi\rangle),
\end{equation} where $|\psi\rangle$ is any purification of $\rho$ and
and $|\varphi\rangle$ for $\sigma$. Then we have
\begin{equation}
T(\rho,\sigma)=\frac{1}{\varepsilon}\cdot\tan{(\arccos{F(\rho,\sigma)})},
\end{equation} where $\varepsilon$ is the precision of the evolution.
\end{Theorem}
{\it Proof.}
\begin{eqnarray}
\aligned
T(\rho,\sigma)=&\min_{|\psi\rangle,|\varphi\rangle}T(|\psi\rangle,|\varphi\rangle)\\
=&\min_{|\psi\rangle,|\varphi\rangle}\frac{1}{\varepsilon}\cdot\tan{(\arccos{F(|\psi\rangle,|\varphi\rangle)})}\\
=&\frac{1}{\varepsilon}\cdot\tan{(\arccos(\max_{|\psi\rangle,|\varphi\rangle}{F(|\psi\rangle,|\varphi\rangle)}))}.
\endaligned
\end{eqnarray}
Applying Uhlmann's theorem \cite{Nielsen00} to the last equation, we
can get
\begin{equation}
T(\rho,\sigma)=\frac{1}{\varepsilon}\cdot\tan{(\arccos{F(\rho,\sigma)})}.
\end{equation}
\hfill $\Box$

In conclusion, we have shown the accurate relation between the
distance of the initial and the final states and the running time of
a class of quantum adiabatic evolution applied in \cite{RC02}. We
have pointed out that via this relation it is convenient to estimate
the running times of some adiabatic algorithms. Furthermore, this
relation can be generalized to the case of mixed states. This
relation maybe can help to design quantum algorithms.

We would like to thank Ji Zhengfeng for useful discussions.

\end{document}